\documentclass[]{aastex63}

\usepackage{graphicx}
\usepackage{advdate}
\usepackage{amsmath}
\usepackage{appendix}

\newcommand{\asec}{\hbox to 1pt{}\rlap{$^{\prime\prime}$}.\hbox to 2pt{}}
\newcommand{\amin}{\hbox to 1pt{}\rlap{$^{\prime}$}.\hbox to 1pt{}}
\newcommand{\adeg}{\hbox to 1pt{}\rlap{$^{\circ}$}.\hbox to 2pt{}}
\newcommand{\efill}{\hskip 1.9pt}
\newcommand{\efild}{\hskip 4.2pt}
\newcommand{\trl}{\color{black}}
\newcommand{\dhm}{\color{black}}

\newcommand{\BV}[1]{\mathbf{#1}}
\newcommand{\BH}[1]{\hat{\mathbf{#1}}}
\newcommand{\BL}{\boldsymbol{\Lambda}}
\newcommand{\CROSS}{\mathbf{\times}}
\newcommand{\DOT}{\boldsymbol{\cdot}}
\newcommand{\gradient}{\boldsymbol{\nabla}}

\submitjournal{{\it The Astronomical Journal}}

\shortauthors{Lauer, Munro, Spencer, et al.}
\shorttitle{Interstellar Navigation}
\begin{document}

\title{A Demonstration of Interstellar Navigation Using New Horizons}

\author[0000-0003-3234-7247]{Tod R. Lauer}
\affil{NSF National Optical Infrared Astronomy Research
Laboratory,\footnote{The NOIRLab is operated by AURA, Inc.
under cooperative agreement with NSF}
P.O. Box 26732, Tucson, AZ 85726; tod.lauer@noirlab.edu}

\author[0000-0002-0662-6923]{David H. Munro}
\affil{Livermore, CA, 94550}

\author[0000-0003-4452-8109]{John R. Spencer}
\affil{Department of Space Studies, Southwest Research Institute, 1301 Walnut St., Suite 300, Boulder, CO 80302}

\author[0000-0003-0854-745X]{Marc W. Buie}
\affil{Department of Space Studies, Southwest Research Institute, 1301 Walnut St., Suite 300, Boulder, CO 80302}

\author[0000-0001-5749-1507]{Edward L. Gomez} 
\affiliation{Las Cumbres Observatory, 6740 Cortona Dr, Suite 102, CA 93117}
\affiliation{Cardiff University, School of Physics and Astronomy, Cardiff, CF24 3AA}

\author[0000-0001-8017-5115]{Gregory S. Hennessy} 
\affiliation{Defense and Mission Support Division US Naval Observatory, 3450 Massachusetts Ave NW, Washington DC, 20392}

\author[0000-0002-9061-2865]{Todd J. Henry}
\affil{RECONS Institute, Chambersburg, PA 17021}

\author[0000-0001-7076-344X]{George H. Kaplan} 
\affiliation{Contractor, U.S. Naval Observatory, 3450 Massachusetts Ave NW, Washington, DC 20392}

\author[0000-0003-0497-2651]{John F. Kielkopf} 
\affiliation{Department of Physics and Astronomy, University of Louisville, Louisville, KY 40292, USA}

\author{Brian H. May}
\affil{London Stereoscopic Company, London, UK}

\author[0000-0002-3672-0603]{Joel W. Parker}
\affil{Department of Space Studies, Southwest Research Institute, 1301 Walnut St., Suite 300, Boulder, CO 80302}

\author[0000-0003-0333-6055]{Simon B. Porter}
\affil{Department of Space Studies, Southwest Research Institute, 1301 Walnut St., Suite 300, Boulder, CO 80302}

\author[0000-0002-1864-6120]{Eliot Halley Vrijmoet}
\affil{Five College Astronomy Department, Smith College, Northampton, MA 01063}
\affil{RECONS Institute, Chambersburg, PA 17021}

\author[0000-0003-0951-7762]{Harold A. Weaver}
\affil{The Johns Hopkins University Applied Physics Laboratory,
Laurel, MD 20723-6099}

\author[0000-0002-4644-0306]{Pontus Brandt}
\affil{The Johns Hopkins University Applied Physics Laboratory,
Laurel, MD 20723-6099}

\author[0000-0003-3045-8445]{Kelsi N. Singer}
\affil{Department of Space Studies, Southwest Research Institute, 1301 Walnut St., Suite 300, Boulder, CO 80302}

\author[0000-0001-5018-7537]{S. Alan Stern}
\affil{Space Sector, Southwest Research Institute, 1301 Walnut St., Suite 300, Boulder, CO 80302}

\author[0000-0002-3323-9304]{Anne. J. Verbiscer}
\affil{University of Virginia, Charlottesville, VA 22904}

\author{Pedro Acosta}
\affil{Asunci\'{o}n, Paraguay}

\author{Nicol\'{a}s Ariel Arias}
\affil{Banfield, Argentina}

\author{Sergio Babino}
\affil{Sociedad Astron\'{o}mica  Octante, Montevideo, Uruguay}  

\author{Gustavo Enrique Ballan}
\affil{Rosario, Argentina}

\author{Víctor \'{A}ngel Buso}
\affil{Observatorio Astron\'{o}mico Busoniano, Santa Fe, Argentina}

\author{Steven J. Conard}
\affil{The Johns Hopkins University Applied Physics Laboratory,
Laurel, MD 20723-6099}

\author{Daniel Das Airas}
\affil{Observatorio Astron\'{o}mico Cielos de Caballito, Capital Federal, Argentina}

\author{Giorgio Di Scala} \affil{Sidney, Australia}

\author{C\'{e}sar Fornari}
\affil{Galileo Galilei X31 Observatory, Oro Verde, Argentina}

\author{Jossiel Fraire}
\affil{Torre\'{o}n, México}

\author{Brian Nicol\'{a}s Gerard}
\affil{Observatorio Tortuga, Col\'{o}n,  Argentina}

\author{Federico Gonz\'{a}lez}
\affil{Rocha,  Uruguay}

\author{Gerardo Goytea}
\affil{Casiilda, Argentina}

\author{Emilio Mora Guzm\'{a}n}
\affil{San Rafael de Santa Cruz, Turrialba, Costa Rica}

\author{William Hanna}
\affil{Columbia Falls, MT 59912}

\author[0000-0002-6131-9539]{William C. Keel}
\affil{University of Alabama, Tuscaloosa, AL 35487-0324,}

\author{Aldo Kleiman}
\affil{Rosario, Argentina}

\author{Anselmo L\'{o}pez}
\affil{Observatorio El Joyero, Astroamigos Col\'{o}n, Col\'{o}n, Argentina}

\author{Jorge Gerardo Machuca} \affil{Torre\'{o}n, M\'{e}xico}

\author{Leonardo M\'{a}laga} \affil{Observaci\'{o}n Astron\'{o}mica Mar del Plata, Mar del Plata, Argentina}

\author{Claudio Mart\'{i}nez} \affil{Fundaci\'{o}n AZARA, Instituto Latinoamericano de Astroturismo, Buenos Aires,  Argentina}

\author{Denis Martinez} \affil{Obsevatorio Tharsis, Las Grutas, Argentina}

\author{Ra\'{u}l Meli\'{a}} \affil{G\'{a}lvez,  Argentina}

\author{Marcelo Mon\'{o}poli} \affil{Asociaci\'{o}n Argentina Amigos de la Astronomía, Capital Federal, Argentina}

\author{Marc A. Murison}
\affil{U.S. Naval Observatory Flagstaff Station
10391 Naval Observatory Road, Flagstaff, AZ 86005}

\author{Leandro Emiliano Fernandez Pohle} \affil{La Rioja,  Argentina}

\author{Mariano Ribas}  \affil{Planetario Galileo Galilei, Capital Federal, Argentina}

\author{Jos\'{e} Luis Ram\'{o}n S\'{a}nchez} \affil{Observatorio  Astron\'{o}mico Géminis Austral, Santa Fe, Argentina} 

\author{Sergio Scauso} \affil{Observatorio B612, Villa María, Argentina} 

\author{Dirk Terrell}
\affil{Department of Space Studies, Southwest Research Institute, 1301 Walnut St., Suite 300, Boulder, CO 80302}

\author{Thomas Traub}
\affil{Martz-Kohl Observatory, Martz Memorial Astronomical Association Inc., Frewsburg, NY 14738}

\author{Pedro Oscar Valenti} \affil{Observatorio Astron\'{o}mico Guillermo Spagnuolo y Grupo Austral de Observadores de Meteoros, Buenos Aires, Argentina}

\author{\'{A}ngel Valenzuela} \affil{ Ituzaingo, Argentina}

\author[0000-0002-5775-2866]{Ted von Hippel}
\affil{Embry Riddle Observatory, Embry Riddle Aeronautical University, Daytona Beach, FL 32114}

\author[0000-0003-0262-272X]{Wen Ping Chen}
\affil{Institute of Astronomy, National Central University, Jhongli, 32001 Taoyuan, Taiwan}

\author{Dennis Zambelis}
\affil{Newcastle 2303, NSW Australia}

\begin{abstract}

As NASA's New Horizons spacecraft exits the Solar System bound for interstellar space, it has traveled so far that the nearest stars have shifted markedly from their positions seen from Earth.  We demonstrated this by imaging the Proxima Centauri and Wolf 359 fields from Earth and New Horizons on 2020 April 23, when the spacecraft was 47.1 au distant. The observed parallaxes for Proxima Centauri and Wolf 359 are $32\asec4$ and $15\asec7,$ respectively. These measurements are not of research grade, but directly seeing large stellar parallaxes between two widely separated simultaneous observers is vividly educational. Using the New Horizons positions of the two stars alone, referenced to the three-dimensional model of the solar neighborhood constructed from Gaia DR3 astrometry, further provides the spacecraft spatial position relative to nearby stars with 0.44 au accuracy. The range to New Horizons from the Solar System barycenter is recovered to 0.27 au accuracy, and its angular direction to $0\adeg4$ accuracy, when compared to the precise values from NASA Deep Space Network tracking. This is the first time optical stellar astrometry has been used to determine the three-dimensional location of a spacecraft with respect to nearby stars, and the first time any method of interstellar navigation has been demonstrated for a spacecraft on an interstellar trajectory. We conclude that the best astrometric approach to navigating spacecraft on their departures to interstellar space is to use a single pair of the closest stars as references, rather than a large sample of more distant stars.

\end{abstract}

\section{Leaving Earth for the Stars}

 We have used the stars to guide our travels on land, at sea, in the air, and in space.  For navigation we largely forego the stellar astrophysics that we might know, reducing the stars to fixed points on the celestial sphere.  Observations of stars with known right ascensions and declinations allow us to find our way on the Earth with longitude and latitude.  For spacecraft, the stars can provide the orientation and direction needed to maintain attitude control. Ranging conducted by the NASA Deep Space Network (DSN), or other similar facilities, provides highly accurate physical locations of the spacecraft, but in standard navigation methodologies these measures remain tethered to the Earth.
 
 The stars are not really fixed, of course. For precise navigation one has to worry about the stars' proper motions, parallaxes, and so on; however, these terms can be regarded as small secular effects that can be applied as corrections to the nominal stellar positions provided for a specific epoch.  But when we leave the Earth and the Solar System behind to venture out into interstellar space, our trek will induce apparent reflex motions in the angular positions of the stars. The task of finding our way will then be one of {\it interstellar} navigation. Besides needing to know what direction we're headed in, we will also need to know how {\it far} we've traveled.  Our motion will appear to shift the positions of nearby stars with respect to more distant ones --- and that will tell us how far we've gone. We can demonstrate this with images obtained by NASA's New Horizons spacecraft obtained on 2020 April 22-23, when it was then 47 au distant from the Sun, passing through the Kuiper Belt.
 \subsection{New Horizons}

 The New Horizons probe is the fifth robotic spacecraft to leave Earth ultimately bound for interstellar space.  New Horizons' primary mission was to conduct the first exploration of Pluto, with a flyby of that planet on 2015 July 14 when it was 33 au distant from the Sun \citep{pluto}.  This was followed by a flyby of the KBO (Kuiper belt object) Arrokoth on 2019 January 1 at 44 au from the Sun \citep{arrokoth}. With an asymptotic outward velocity of 14 km/s, each year it becomes $\sim 3$ au more distant. At this writing, New Horizons is 61 au distant, and is in excellent condition, with good prospects for remaining scientifically useful at least out to 100 au. The spacecraft is capable of conducting a close flyby of a second KBO, should a suitable target be located. It is now conducting a program of distant KBO observations and heliophysics.  A year ago, New Horizons conducted a program of astrophysical observations, which included a measuring the intensity of the cosmic optical background \citep{postman}.
 
 In evaluating what astrophysical observations might be worthwhile from the unique location of New Horizons, the possibility of using its large Earth-spacecraft (hereafter ES) baseline for stellar-parallax\footnote{In this narrative we use two somewhat different context-dependent definitions of the word ``parallax.'' In general, we take parallax to mean the observed shift or ``parallactic displacement" of an object against a distant background when viewed from two different locations. This is distinct from {\it the} standard parallax of a star as seen with a 1 au baseline.} measurements was often raised.  However, it was readily apparent from the known performance of the New Horizons instrumentation \citep{lorri, lorri2} that it could not produce astrometric measurements even remotely competitive with Gaia and other ongoing astrometric programs, despite its large baseline. At the same time, with coordinated Earth-based observations, New Horizons could provide easily visualized parallactic observations, which would have pedagogical value  for demonstrating the concept of stellar parallaxes.  Likewise, such observations could also provide direct demonstration of the great distance that the spacecraft had traveled.  Further, with two or more astrometric observations of the nearest stars, a novel demonstration of autonomous interstellar navigation, using New Horizons' cameras alone, could be conducted.  As such, on 2020 April 22-23 we used New Horizons' LORRI (Long-range Reconnaissance Imager) camera to image the star fields containing the nearby stars Proxima Centauri and Wolf 359. We present the details of this observational program, the results from the stellar parallax measurements, and the interstellar navigation demonstration, in the sections that follow.
 To enhance the educational value of this program, we have deposited a Jupyter notebook plus the images used in the analysis to Zenodo: doi:10.5281/zenodo.15359866.  Additional images and data products are available from APL at https://pluto.jhuapl.edu/Learn/Parallax/Parallax-Images.php.

 \section{The New Horizons Parallax Program}\label{sec:parallax}

 When teaching the concept of stellar parallaxes in the classroom, a common approach is to ask the students to hold up a finger against a more distant background, and look at it with one eye at a time.  The parallactic displacement is obvious, as the projection of the finger against the background viewed from the left versus right eye jumps back and forth. Unfortunately, observation of stellar parallaxes is considerably more involved.  While the Earth moving around to a diametrically-opposed point in its orbit after six months can be described as defining a baseline analogous to the separation of our two eyes, stellar proper motions are typically similar in magnitude to the parallax, itself.  Real parallax measurements require observing the star for a number of years, with the parallax evident as a periodic wobble superimposed on the linear drift caused by the star's proper motion.  And there isn't much to see directly.  Even the nearest stars have parallaxes less than an arcsecond, which are typically much smaller than the PSFs of ground based telescopes, and all but a few space telescopes. While accurate astrometry can be obtained to an arbitrarily small fraction of the PSF width, the parallactic displacements of a star are not evident by casual visual inspection of an image. The reflex motion of a star due to parallax will only show up as a nearly imperceptible wiggling in the sequence of images used to derive the parallax. In professional astrometric programs, the parallaxes indeed are only realized as parameters recovered from careful numerical analysis of several years of observations.

 In contrast, the shifts in position measured with a simultaneous Earth/New Horizons observations are a significant fraction of an arcminute for the nearest stars.  No proper motion corrections are needed in this particular case.\footnote{Proper motion corrections, however, will be required for the navigation demonstration presented in the next section.} Direct inspection of the Earth-based versus New Horizons images instantly reveals the parallactic displacement of the target star with respect to background stars. In effect, this approach is the true analogue to the exercise of holding a finger in front of you and blinking your eyes.

 \subsection{Selection of Program Stars}

 Definition of the New Horizons parallax demonstration program began in 2019 after completion of the Arrokoth encounter, with the intent of conducting the program in 2020, when the associated post-encounter spacecraft operations would be completed.  One goal was to engage the communities of both professional and amateur astronomers in the task of observing the target stars simultaneously with the New Horizons imaging observations. We decided to observe two stars to provide for a demonstration of interstellar navigation. Selecting two stars apart in angle on the sky by $\sim90^\circ$ would provide for the best recovery of the spacecraft position.  For operational simplicity, we wanted to conduct all the observations in a single command load, thus the stars would be observed within a few days of each other.

 The size of the parallactic displacement for a given star depends not only on the distance to the star, but also the projection of the ES baseline onto the plane of sky at the star's position. The displacement is maximized for stars sighted perpendicular to the baseline, To develop the sample, we searched through the 50 stars nearest to the Sun to identify the stars having the largest dispalcements, given the orientation of the ES baseline.  One restriction was that the brightest nearby stars, like $\alpha$ Centauri or Sirius, would be over-exposed in exposures deep enough to capture the fainter background reference stars. Another concern was to avoid strong scattered sunlight in the LORRI camera, so stars within a few dozen degrees of the Sun were not considered as well\footnote{Since New Horizons is exiting the Solar System on an almost radial trajectory, the Sun's celestial coordinates as observed by the spacecraft are essentially constant. Stars now angularly close to the Sun as seen from New Horizons will remain so.}.

 Fortuitously, Proxima Centauri, the nearest star beyond the Solar System, had the largest ES displacement and was easily observable by New Horizons. Proxima Cen would provide the most dramatic demonstration of the large parallactic displacements that could be observed with New Horizons, and thus was a compelling target.  Proxima Cen was best observed from Earth in April, so the program was designed to run over 2020 April 22-23, close to the time of new moon to provide dark skies for Earth-based observers. At that epoch, Proxima Cen had an ES parallax of $\sim32''.$

 The one disadvantage to observing Proxima Cen was its declination of $-62^\circ,$ making it inaccessible to nearly all observers in the northern hemisphere.  Fortunately, the star with the third largest ES displacement, Wolf 359, was well placed for northern hemisphere observation in April, having a declination of $+7^\circ$ and a right ascension only 3.5h earlier than that for Proxima Cen\footnote{UV Ceti by a slight margin had the second largest displacement, but was at $-17^\circ$ declination, and could only be observed from Earth several months later.}. In 2020 April, Wolf 359 had an ES displacement of $\sim16''.$ In passing, we note that Wolf 359 has been featured in more than one science fiction story\footnote{While Wolf 359 is most commonly known from its appearance in the {\it Star Trek Next Generation} TV series, it was also featured in the ``Wolf 359" episode of the 1964-65 TV series {\it The Outer Limits.}} and was thus well known outside of the professional astronomical community.

 \begin{deluxetable}{lccl}
\tabletypesize{\small}
\tablecolumns{4}
\tablewidth{0pt}
\tablecaption{Target Star Properties}
\tablehead{
\multicolumn{1}{l}{} &
\multicolumn{1}{c}{Proxima Centauri} &
\multicolumn{1}{c}{Wolf 359} & 
\multicolumn{1}{l}{Reference}
}
\startdata
Type & M5.5Ve & dM6 & \citet{bessell, kesseli} \\
$m_V$&11.13 & 13.507 & \citet{jao, landolt} \\
Parallax & $0\asec768066(50)$ &$0\asec415179(68)$ & \citet{gaia} \\
Distance (pc) & $1.301971(85)$ &$2.40860(40)$ & Reciprocal of parallax\\
$\alpha$ proper motion (mas/yr) & $-3781.741(31)$ & $-3866.338(81)$  & \citet{gaia} \\
$\delta$ proper motion (mas/yr) & $\phantom{-3}769.465(51)$&$-2699.215(69)$&\citet{gaia} \\
Barycentric radial velocity (km/s) & $-20.57820$& 19.57 & \citet{apogee, fouque} \\
\enddata
\tablecomments{In this and other tables we provide the standard deviations of the stated parameters in shorthand notation. The uncertainties are given in parentheses at the end of the parameter value, with the digits given corresponding the final digits in the parameter.}
\end{deluxetable}\label{tab:stars}

The astrophysical parameters of Proxima Cen and Wolf 359 are listed in Table \ref{tab:stars} and their celestial coordinates are given in Table \ref{tab:coords}. The present analysis is anchored by the high-accuracy coordinates provided by the Gaia DR3 catalog \citep{gaia}. The origin of the Gaia coordinate system is defined to be the Solar System barycenter, with the epoch set to 2016.0.  The proper motions of both Proxima Cen and Wolf 359 are substantial, thus proper motion corrections to the New Horizons epochs of observation were required  for the navigation exercise, as well as to optimally center both stars in the LORRI aperture. The proper motion vectors for both stars are provided by the Gaia DR3 catalog and are given in Table \ref{tab:stars}.  Using the observational epochs provided in Table \ref{tab:obs}, the proper motion corrections were applied as a small rotation to the epoch 2016.0 coordinates.

At the level of Gaia's accuracy, the integrated {\it radial} motions of the target stars over the interval of time between the DR3 and New Horizons epochs will also begin to be detectable, given the stars' radial velocities.  Our transformation of stellar coordinates between the two epochs uses the full three-dimensional motion of the stars derived from the combination of their proper motions and radial velocities. In the present case, however, the New Horizons astrometry is not precise enough to be sensitive to the small changes in distances to the stars between the two epochs, so change in radial distance between the two epochs makes no discernible difference for our analysis. The final coordinates corrected to the New Horizons epoch are given in Table \ref{tab:coords} on the line below the Gaia DR3 coordinates.

\begin{deluxetable}{lllll}
\tabletypesize{\small}
\tablecolumns{5}
\tablewidth{0pt}
\tablecaption{ICRF Stellar Coordinates}
\tablehead{
\multicolumn{1}{l}{} &
\multicolumn{2}{c}{Proxima Centauri} &
\multicolumn{2}{c}{Wolf 359} \\
\multicolumn{1}{l}{Parameter} &
\multicolumn{1}{c}{$\alpha$} &
\multicolumn{1}{c}{$\delta$} &
\multicolumn{1}{c}{$\alpha$} &
\multicolumn{1}{c}{$\delta$}
}
\startdata
Gaia DR3 (epoch 2016.0)& 217.39232147 & $-62.67607512$ &164.10319031& $+7.00272694$\\
 Gaia DR3 pm corrected to NH epoch &217.38246430 & $-62.67515412$ & 164.09852751& $+6.99949593$ \\
 Observed from New Horizons&217.36315(11)&$-62.676324(16)$&164.0943006(69)&$+7.001008(11)$ \\
 Gaia$+$JPL estimate from New Horizons&217.36311984& $-62.67633603$& 164.09426343&$+7.00103475$ \\
 Observed from Earth at NH epoch & 217.3826035(47) &  $-62.6752941(22)$ & 164.0984407(11) &$+6.9995292(11)$\\
 Gaia$+$JPL estimate from Earth at NH epoch&217.38252116& $-62.67531791$&164.09844471& $+6.99953024$
 \enddata
 \tablecomments{All parameters are given in degrees. The ``Gaia$+$JPL" estimates are based on Gaia DR3 positions of the target stars combined with the JPL Horizons knowledge of the spacecraft and Earth positions relative to the Solar System barycenter.  The New Horizons and Earth-based coordinates have been implicitly corrected for stellar aberration. As such, they correspond to stellar positions measured by observers at rest with respect to the SSB at the locations of Earth and New Horizons. The visit 1 and 2 average distance of New Horizons from the Solar System barycenter is 47.1176 au; the distance of the Earth from the barycenter for the same epoch is  1.00563 au.}
\end{deluxetable}\label{tab:coords}

\subsection{The Astrometric System, and Treatment of Stellar Aberration}

In basing our program on Gaia DR3 astrometry, we are expressing our celestial coordinates and associated analysis in the International Celestial Reference Frame (ICRF).  In short, the origin of ICRF coordinates is the solar system barycenter (SSB) at rest.  As such we are not concerned with the Earth-based effects of precession, nutation, and so on. Significantly, ICRF stellar coordinates are corrected to zero parallax.

Both the Earth and New Horizons are in motion, of course, and {\it absolute} coordinates measured at either station will be affected by ``stellar" or ``velocity" aberration.  This is the classic effect in which the apparent position of a star will be displaced by several arcseconds from its ``at rest" position by motion of the observer.  For the case of a non-relativistic observer velocity, $v\ll c,$ where $c$ is the speed of light, the observed angle of the star, $\phi,$ relative to the observer's velocity vector is:
\begin{equation}\label{eq:vab}
    \phi=\arctan\left(\frac{\sin{\theta}}{\beta+\cos{\theta}}\right),
\end{equation}
where $\beta=v/c,$ and $\theta$ is the true angle of the star with respect to the velocity vector.  Note that $\phi<\theta;$ the observer has to lean over in the direction of  their motion.  For both the motion of Earth and New Horizons the apparent displacement of the star $\theta-\phi,$ can be several arcseconds.  For Proxima Cen and Wolf 359 on 2020 April 23 the velocity of the spacecraft relative to the SSB was  13.97 km/s, which gave stellar aberration offsets of $8\asec8$ and $7\asec3$ for Proxima Cen ($\theta=66\adeg8$) and Wolf 359 ($\theta=130\adeg2$), respectively.  For the Earth with its orbital velocity of $\sim30$ km/s, the stellar aberration can produce offsets roughly twice as big.

In practice, however, we can ignore the stellar aberration since we only measure the positions of the target stars relative to the background stars present in the small fields surrounding them.\footnote{As noted in \citet{isn_3}, if we did know the absolute direction the camera was pointed with arcsecond accuracy, we could use the aberration of the background star fields relative to their ICRF directions to infer the velocity of the spacecraft relative to the SSB. Stellar aberration corrections are in fact required to interpret the orientation telemetry provided by the spacecraft's star trackers.} The amplitude of the offset as given by eqn (\ref{eq:vab}) varies slowly with $\theta,$ making differential changes in $\theta-\phi$ over the camera fields negligible when compared to the present astrometric errors. For example, for the NH observations of Proxima Cen, the $\theta-\phi$ offset going from the center of the field to its edges decreases by only  $0\asec010$.  If the first-order effect of stellar aberration is to shift the positions of all the stars in the field by the same amount, then the second-order effect is a small isotropic change in the pixel-scale of the images.  For $\theta<90^\circ$ stellar aberration will cause the angular field of the camera to shrink very slightly (for New Horizons), while it will expand slightly for $\theta>90^\circ.$

In other words, using the ICRF coordinates of the background stars directly still produces highly accurate astrometry of the target stars. With the target stars positioned in the center of the fields, and a small allowance in the astrometric solution made for a variable pixel scale, the astrometry will be even more accurate. The only caveat, as we discuss in the notes to Table \ref{tab:coords}, is that the ICRF coordinates reported for the Earth and New Horizons-based astrometry technically correspond to observers stationary with respect to the SSB at the positions of the Earth and New Horizons.

\subsection{New Horizons Observations}\label{sec:nhobs}

The parallax images were obtained with New Horizons' LORRI (Long Range Reconnaissance Imager) camera. LORRI obtained the highest resolution images of Pluto and Arrokoth, but was also used to provide the terminal navigation to both objects.  Significant to the present work, navigation to Arrokoth was especially challenging, and accurate astrometry obtained with LORRI was critical for the success of that encounter \citep{lorri_nav}. Extensive discussions of the instrument are provided by \citet{lorri} and \citet{lorri2}, but briefly, LORRI is an unfiltered (white light) $1024\times1024$ pixel CCD imager mounted on a 20.9 cm aperture Cassegrain reflector. LORRI's passband extends from $0.4~\mu{\rm m}$ $0.9~\mu{\rm m},$ with a pivot wavelength of 0.608$~\mu$m. 

For the most sensitive astronomical imaging, the camera is operated with $4\times4$ pixel binning, producing raw images in $256\times256$ pixel format. This mode was used for the present observations. The pixel-scale in $4\times 4$ mode is $4\asec08,$ which provides a $17\amin4$ field. The PSF in this mode is highly under-sampled, with the FWHM markedly less than two pixels. Guiding is provided by the spacecraft attitude control thrusters, which can cause the detailed shape of the PSF to vary somewhat from exposure to exposure. In $4\times 4$ mode the gain is $19.4e^-$/DN, and the read-noise is $24e^-.$ The photometric zeropoint is $18.88\pm0.01$ AB magnitudes, corresponding to a 1 DN/s exposure level.

Both stars were observed twice by New Horizons within a 28 h interval spanning 2020 April 22-23.  As both stars are flare stars and might have been over-exposed by LORRI, two visits provided for some contingency.  This strategy also allowed for improved coverage to guard against bad weather at Earth-based sites. A small complexity was to plan the timing of the observations to arrange for simultaneous Earth and spacecraft observations of the two stars. The spacecraft clock is synchronized to universal time (UT), but at the time of the observations, with the ES separation of 46.85 au, the light travel time between the two stations was 6.49 h, allowing various ways to define ``simultaneity."  Our choice was to specify the timing so that both stations would observe the same ``events" on the target stars.  Given the directions of the stars on the sky and the orientation of the ES baseline vector, this meant that New Horizons would {\it lead} the Earth observations by 2.88 h for Proxima Cen, but {\it follow} the Earth observations of Wolf 359 by 3.74 h.

Eight LORRI images were obtained at the first visit for each star. Three exposures were made at the nominal best estimate of the mid-range exposure time, with two more exposures at 10\%\ of that time, and three at $10\times$ the nominal exposure time to bring up fainter stars in the field.  The multiple exposures at a given exposure time allowed for the detection and repair of cosmic-ray hits, which are substantial for LORRI, while the shortest exposures allowed for modest flares and under-estimation of the signal level in the nominal exposures. Five images were obtained at the second visits, which repeated only the nominal and shortest exposures.  The set of observations is listed in Table \ref{tab:obs}. The reduced LORRI images are avaiable from the NASA Planetary Data System archive (https://pds.nasa.gov/ ), and can be located by the MET designations given in Table \ref{tab:obs}.

\subsubsection{New Horizons Astrometry}

Astrometric measurements of the positions of Proxima Cen and Wolf 359 were obtained from the three mid-range exposures obtained during the two visits for each star. While the spacecraft telemetry provides pointing information accurate to the sub-arcsecond level, deriving the pointing from the Gaia DR3 coordinates of background stars in the images provided the most accurate results. The astrometric calibration and measurement of the target star positions were done separately for each of the six images.

In detail, stars were detected in each image and matched to entries in the Gaia catalog, with proper motion corrections applied to transform the ICRF coordinates to the New Horizons epoch.  Proxima Cen was surrounded by a rich star field, and $\sim230$ Gaia stars were measured in each of its three images. The Wolf 359 fields were markedly sparser, with only 14 to 18 reference stars available.  Given the different exposure times used for the two stars (0.5 s for Proxima Cen vs. 5 s for Wolf 359), the photometric depth was also markedly different for the two stars. Both stars were the brightest source in their LORRI fields. The reference stars extended to sources $\sim 4.5$ mag fainter.

We note that measuring stellar astrometry from LORRI $4\times4$ images presents some challenges, given the severe under-sampling of the PSF plus the strong flux of cosmic rays events in even short LORRI exposures.  The multiple exposures for each star are used to recognize cosmic rays hits, but under-sampling also means that the images cannot be combined at the sub-pixel level without incurring significant degradation of the resolution \citep{psf}. 

We concluded that differential chromatic aberration, which might have been of concern given the strong red color of both target stars, should not significantly affect the astrometry. The LORRI telescope is a reflector.  It does have fused silica field-flattening lenses \citep{lorri}, but they have low power, and the target stars are centered on the optical axis.  In practice, highly accurate LORRI astrometry was obtained of Arrokoth to support the terminal navigation to the object \citep{lorri_nav}, despite that it is also strongly red \citep{arrokoth}.

A separate question is if any background stars might have parallaxes that need to be accounted for. Based on the Gaia DR3 catalog, we see that even with the large NH baseline, nearly all background stars in both fields have ES parallaxes less than $0\asec1$, well within the $0\asec18$ error that we will assign to the final target-star coordinates (see $\S\ref{sec:errors}$). That said for high-precision navigation, the displacements of the background stars are correlated with the displacement of the target star, creating a small systematic effect that can be corrected for; we ignore this term in the present exercise. Lastly, we identify one star in the Wolf 359 field  (Gaia DR3 3864968128241533056; $\alpha=164\adeg189883,~\delta=6\adeg900660$) predicted to have an ES parallax of $0\asec79,$ which would be detectable in the LORRI images.  

The final coordinates for the two target stars are given in Table \ref{tab:coords} on the ``Observed from New Horizons" line, and are the average over the six mid-range obtained for each. The errors quoted in Table \ref{tab:coords} are the statistical errors in the mean values.  

\begin{deluxetable}{lcllcll}
\tabletypesize{\small}
\tablecolumns{7}
\tablewidth{0pt}
\tablecaption{LORRI Observations}
\tablehead{
\multicolumn{1}{l}{Star} &
\multicolumn{1}{c}{Visit} &
\multicolumn{1}{c}{Earth (UT)} &
\multicolumn{1}{c}{NH (JD)} &
\multicolumn{1}{c}{Exposures (s)} &
\multicolumn{1}{c}{MET Start} &
\multicolumn{1}{c}{MET End}}
\startdata
Proxima Cen & 1 & 2020 April 22 13:00& 2458961.9215 &$2\times0.05,~3\times0.5,~3\times5$&0449855930&0449855974\\
Proxima Cen & 2 & 2020 April 23 05:00& 2458962.5881 &$2\times0.05,~3\times0.5$& 0449913501&0449913533 \\
Wolf 359& 1 & 2020 April 23 04:00& 2458962.8231& $2\times0.5,~3\times5,~3\times50$&0449933793&0449933982 \\
Wolf 359 & 2 & 2020 April 23 10:00& 2458963.0734 &$2\times0.5,~3\times5$&0449955392&0449955466
 \enddata
\tablecomments{The Earth UT times are when an Earth-based observer would see the same stellar events as the New Horizons executing observations at the specified Julian date. MET is the mission elapsed time designator for the LORRI images in each visit.}
\end{deluxetable}\label{tab:obs}
\subsection{Earth-based Observations}\label{sec:earth}

After we selected 2020 April 22-23 as the epoch of New Horizons observations, we solicited the Earth-based observations through a wide variety of channels, and received expressions of interest from several professional and amateur astronomers.  Unfortunately, in 2020 March the onset of the COVID-19 pandemic caused most professional observatories to shut down operations and severely curtailed the ability of many of the amateur astronomers to participate. In the end the best matches to the passband and timing of the LORRI observations were provided by two professional, remotely operated, telescopes.

The Proxima Centauri image was obtained on April 22 at 12:51 UT with CMOS camera mounted on the  0.4-meter telescope at the Siding Spring, Australia node of the Las Cumbres Observatory. The time of observation was only nine minutes ahead of the nominal first-visit time (see Table \ref{tab:obs}). The exposure was 30s in a Pan-STARRS w filter. With a pivot wavelength of 6250\AA\ and wide passband of 4416\AA, it strongly resembles the LORRI passband. The pixel scale is $0\asec58,$ which provides excellent sampling.

The Wolf 359 image was obtained on April 23 at 04:37 UT with the University of Louisville 0.6 meter Manner Telescope (ULMT)  located at Mt. Lemmon Observatory, near Tucson,  Arizona. The observation trailed the nominal first-visit time by 37 minutes with an exposure of 120 seconds in a Sloan r' filter.  A $4096\times4096$ SBIG STX-16803 camera provided an image scale of $0\asec39$/pixel with a stellar point-spread-function full-width-half-maximum of $1\asec9$. A dark-subtracted $3000\times3000$ pixel region covering $19\amin5 \times 19\amin5$ was provided for subsequent astrometry.

The Earth-based astrometric positions for both stars are shown in Table \ref{tab:coords} on the ``Observed from Earth at NH epoch" line. As with the New Horizons astrometry, the Earth-based astrometry is calibrated by Gaia astrometry of the background stars in the images.

\begin{figure}[htbp]
\centering
\includegraphics[keepaspectratio,width=7.0 in]{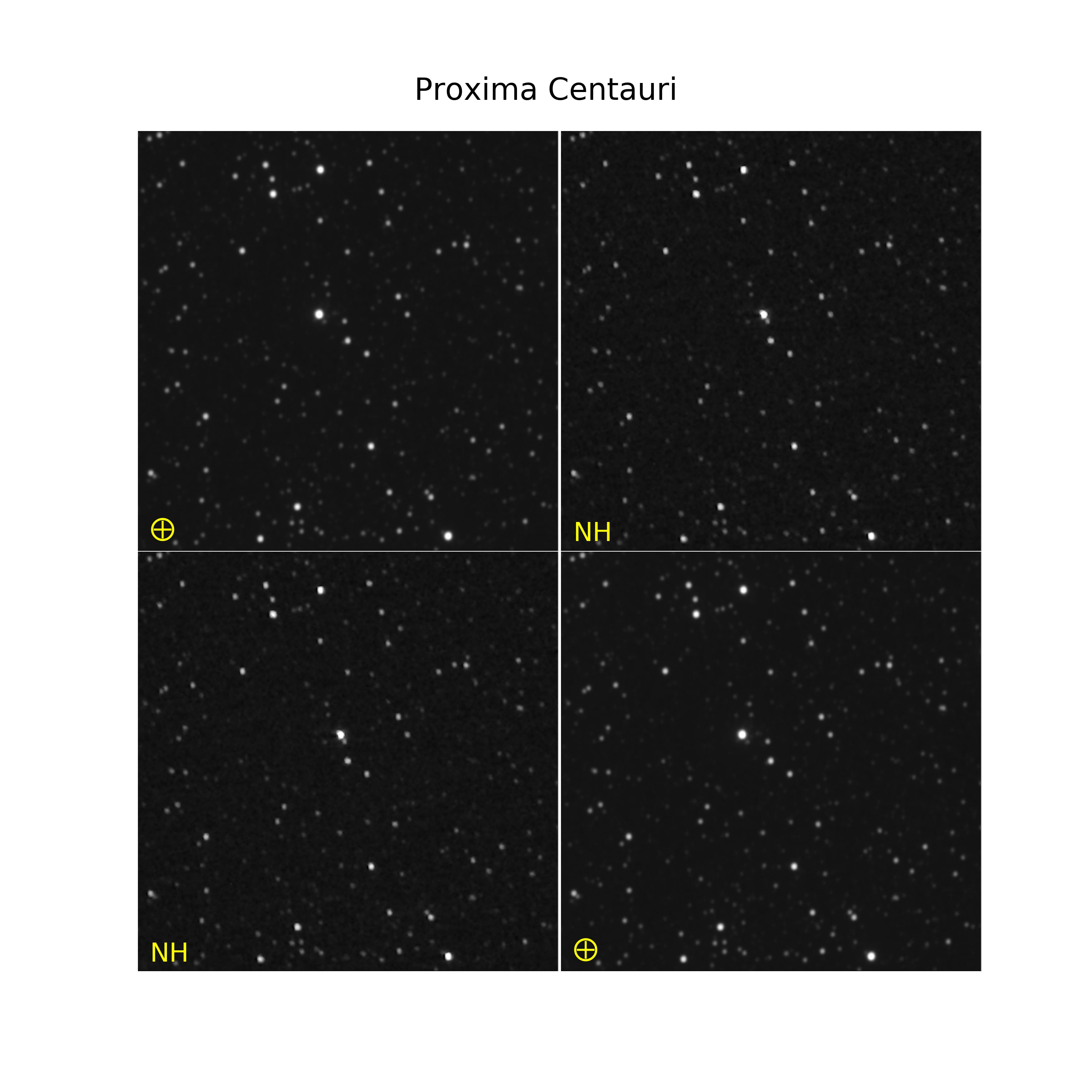}
\caption{The Earth-based and New Horizons images of Proxima Centauri and its star field are shown side by side to demonstrate the large Earth-spacecraft parallax. Proxima Cen is the bright star near the center of the field.  The field shown is $10'\times10'$.  North is at the top. The image pairs have been prepared to a common image scale, field, and orientation so that the parallax can also be recognized with stereo imaging. The top pair is positioned for ``cross-eyed" viewing. Crossing your eyes to view the NH-based image with the left eye, and the Earth-based image with the right eye, will create the appearance of Proxima Cen floating in front of the background stars. The two images are swapped in position in the bottom row to allow for parallel viewing. In this case, the left eye views the left panel, and the right eye the right panel. Parallel viewing can also be done by mounting the images in a stereoscopic viewer. Our experience on the New Horizons team is that there is no clear preference between cross-eyed vs. parallel viewing.}
\label{fig:proxima}
\end{figure}

\begin{figure}[hbtp]
\centering
\includegraphics[keepaspectratio,width=7.0 in]{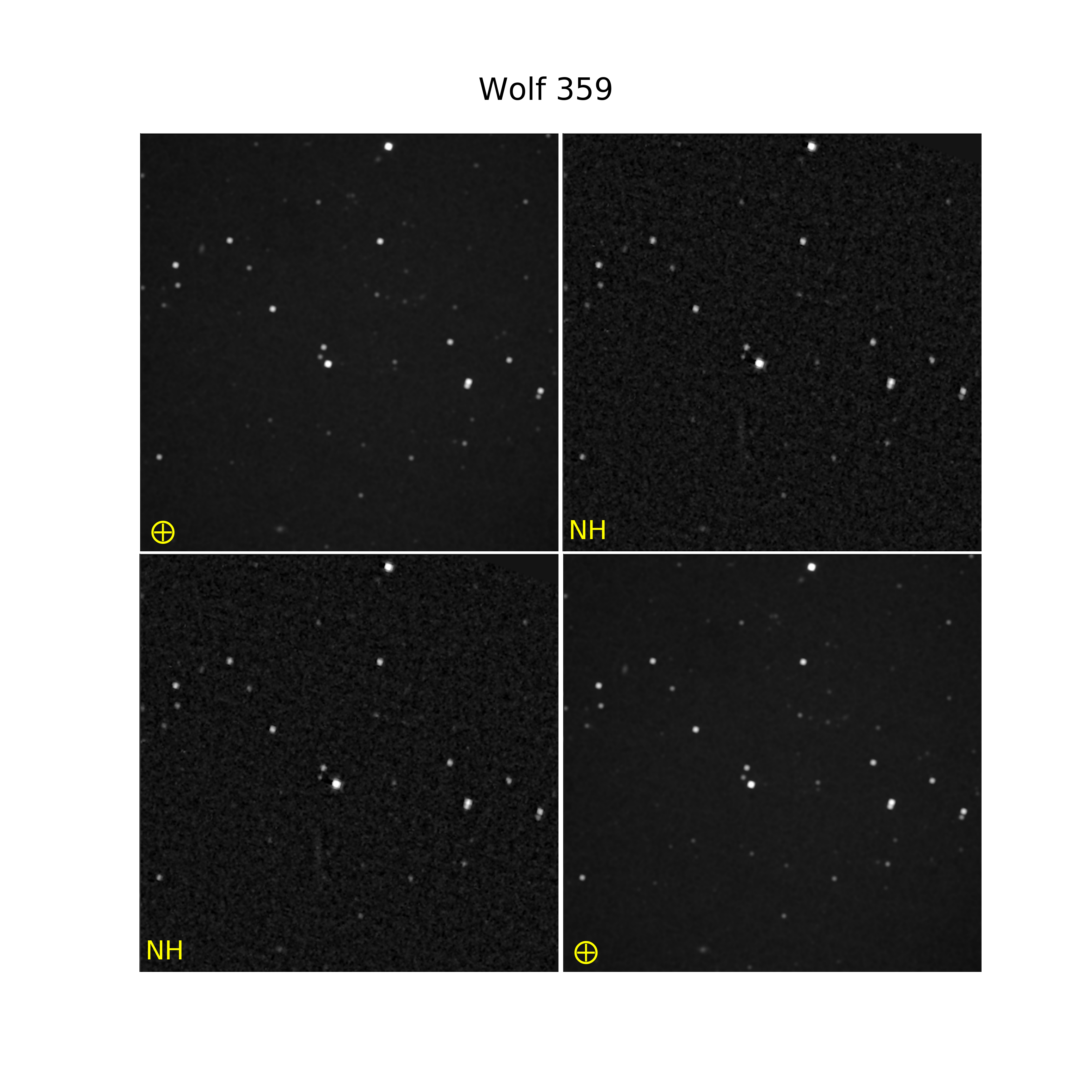}
\caption{As in Figure \ref{fig:proxima}, but for Wolf 359.}
\label{fig:wolf}
\end{figure}
\subsection{Estimation of the Earth-Spacecraft Parallaxes}

The central portions of the New Horizons and Earth-based images are shown in Figure \ref{fig:proxima} for Proxima Cen, and Figure \ref{fig:wolf} for Wolf 359. In these renditions the images have been prepared  to allow stereo imaging to be used to directly visualize the parallactic displacements between the two vantage points. To accomplish this, the ground-based and New Horizons images had to be matched in both scale, orientation, point-spread function, and brightness.   The individual New Horizons images after initial reduction by the standard LORRI pipeline were then given precise World Coordinate System (WCS) headers using automated routines that compared star positions in the images to those from a standard catalog.  Hot pixels and “jailbar” artifacts \citep{lorri2} were removed, then the images in each set were registered using the WCS information, and robustly stacked to improve SNR.   The stacked New Horizons images were enlarged and rotated to match the plate scale and orientation of the Earth-based images, and the Earth-based images were convolved with a Gaussian PSF to approximately match the coarser point-spread function of the New Horizons images.  Finally, both images were stretched to the same intensity scale, and rotated so that the parallactic offsets were horizontal, and cropped to remove edge artifacts.

Simple visual comparison of the Earth and New Horizons images for each star shows that the large displacements are obvious. Further, with stereo imaging, the target stars will appear to float in front of the background stars, which are all too far away to have readily visible parallaxes. Figures \ref{fig:proxima} and \ref{fig:wolf} have been formatted to provide for both parallel and cross-eyed viewing (these images and more are available at the link given at the end of the introduction).

The most direct way to quantify the ES parallaxes, $\pi,$ between the Earth and spacecraft stellar coordinates for either star is to compute the dot product between the two vectors:
\begin{equation}
\pi = \arccos{(\BH{d}_n \DOT \BH{d}_e}),
\end{equation}
where $\BH{d}_n$ and $\BH{d}_e$ are the cartesian coordinate unit-vectors. For example, in terms of New Horizons observed right ascension, $\alpha_n,$ and declination, $\delta_n,$
\begin{equation}\label{eqn:dvec}
{\BH{d}_n} = \left(\cos(\alpha_n)\cos(\delta_n), \sin(\alpha_n)\cos(\delta_n), \sin(\delta_n)\right).
\end{equation}
In passing, we note that the arccos-function requires high-numerical precision to return accurate values for extremely small angles; however, the 64-bit precision incorporated into the Python routines used for the present analysis produces accurate results.

The observed parallactic displacement can be compared to the estimated ES parallax, $\pi',$ which can be derived from the known distance, $r_{se},$ and direction vector to the spacecraft, $\BH{s}_e,$ as compared to the known distance to the star, $r_*~$:
\begin{equation}
\pi' = \frac{r_{se}}{r_*}\left(1-(\BH{s}_e \DOT \BH{d}_e)^2\right)^{1/2}.
\end{equation}
The estimated positions of both Proxima Cen and Wolf 359 at the New Horizons epoch are given in Table \ref{tab:coords}.  They are based on the {\trl Gaia} DR3 barycentric positions of the stars and the JPL Horizons ephemerides for Earth and the New Horizons spacecraft. As such, they can be regarded as the correct answers ``in the back of the book."

For Proxima Centauri the observed and estimated ES parallaxes are $32\asec36$ and  $32\asec27,$ respectively, with difference $0\asec10.$ 
For Wolf 359, the same quantities are $15\asec72,$ $15\asec89,$ and $-0\asec17.$
In finer detail, direct comparison of the observed NH-based positions to the estimated NH-based positions shows the $(\alpha,\delta)$ observed$-$estimated residual pairs to be $(0\asec033,0\asec043)$ for Proxima Cen and $(0\asec132,-0\asec097)$ for Wolf 349.  
Likewise, comparison of the observed Earth-based positions to the estimated Earth positions gives the $(\alpha,\delta)$ residual pairs to be $(-0\asec127,-0\asec087)$ for Proxima Cen and $(0\asec016,0\asec005)$ for Wolf 349.  
\section {Interstellar Navigation}

It has long been assumed that interstellar spacecraft could use stars for navigation \citep{isn_1,isn_2}.  More recently, \citet{isn_3} developed a general approach to the use of stars to guide spacecraft traveling at relativistic velocities on long-duration interstellar missions.  Radio pulsars \citep{radio} and X-ray pulsars \citep{xray} have also been proposed for use as navigational references. Indeed, X-ray pulsar observations with spacecraft in Earth-orbit have been used to provide successful proof of concept demonstrations of autonomous spacecraft navigation \citep{insight, sextant}. Work is underway to develop the technology needed to use X-ray pulsars for autonomous navigation over large distances within the Solar System and beyond.

New Horizons has been navigated through encounters with Jupiter, Pluto and Arrokoth, with high accuracy ranging and positional determinations provided by NASA's Deep Space Network (DSN). This was augmented with LORRI imagery for terminal guidance during the Pluto and Arrokoth encounters.  But given its large distance from Earth, we realized that we could also use onboard imaging of star fields to relatively accurately determine its position vector.  To our knowledge, this is the first time onboard stellar astrometry has been used to provide such information.
\subsection{The Navigation Methodology}

\begin{figure}[hbtp]
\centering
\includegraphics[keepaspectratio,width=5.0 in]{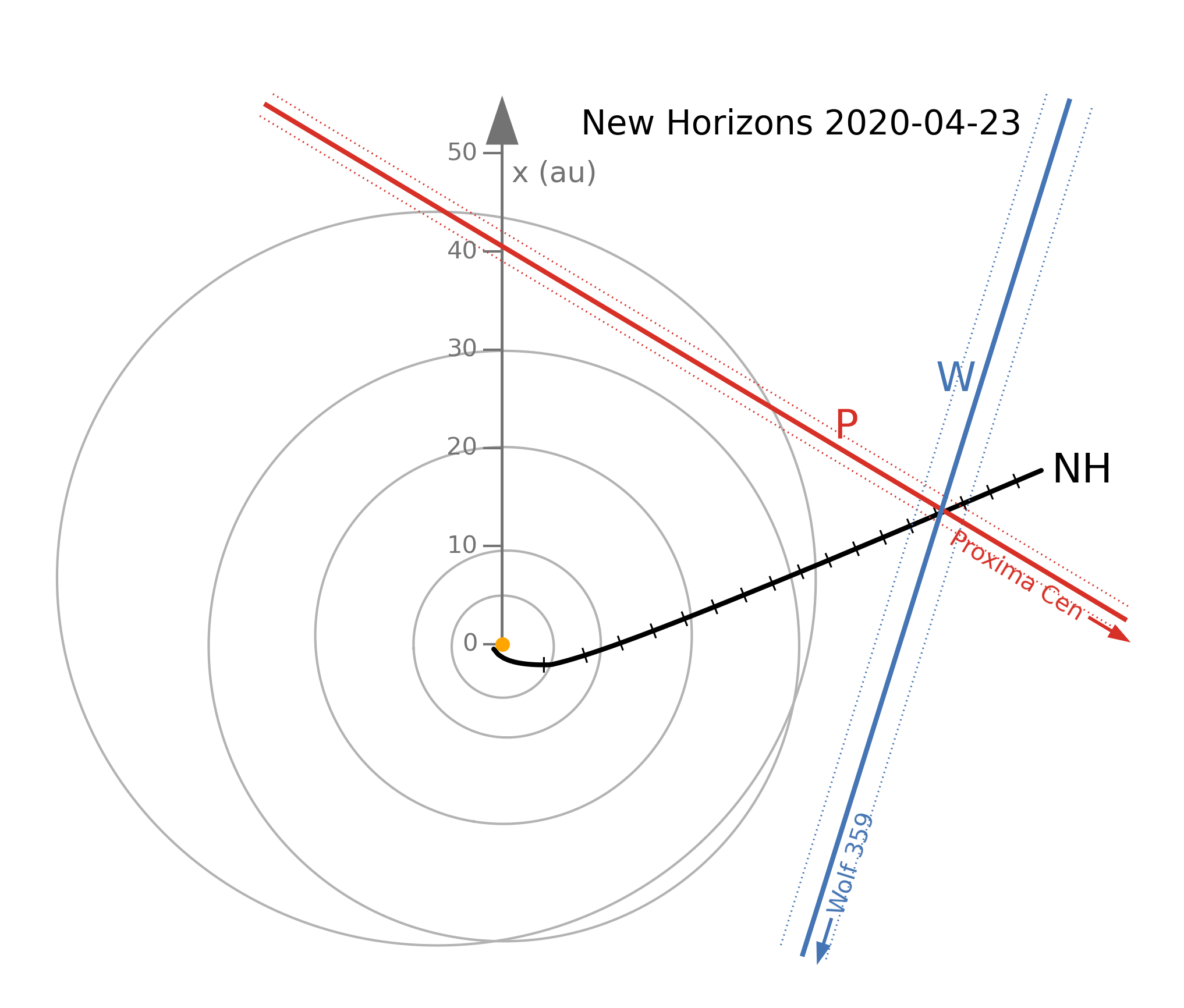}
\caption{The location of New Horizons on 2020 April 23 as derived from the directions to Proxima Cen and Wolf 359 measured from the spacecraft. The view is from the ecliptic north pole; the vertical axis is at zero RA.  Gray circles show the orbits of the outer planets.  Line of position P passes through the Gaia 3-D location of Proxima Cen, in the direction measured from the spacecraft; the observations of Proxima Cen thus constrain the spacecraft to lie on line P. Similarly, observations of Wolf 359 constrain the spacecraft to lie on line of position W.  The faint dotted lines show how much P and W would be displaced by a $1''$ change in line direction; the transverse displacement in au is just the distance to the star in pc (1.30 for P, 2.41 for W).  The trajectory NH is the actual path of the spacecraft from launch in 2006 through 2023, marked with yearly tickmarks.  The actual angular  uncertainties are much less than the $1''$ indicated by the dotted lines.  Line P is inclined $\sim 45^\circ$ from the ecliptic plane; line W and the NH trajectory are inclined less than $2^\circ$ from the ecliptic.}
\label{fig:nav_wide}
\end{figure}

Figure \ref{fig:nav_wide} diagrams the geometry of interstellar navigation using only observations of the apparent directions of stars.  We know with great precision the three-dimensional position vectors $\BV{p}_k$ of nearby stars from the Gaia DR3 catalog.  We measure the apparent directions to these stars, expressed as unit vectors $\BH{d}_k$, from the vantage point of the spacecraft.  Each such measurement constrains the spacecraft to lie on a ``line of position'' (LOP), that is, the line with direction $\BH{d}_k$ passing through the star at point $\BV{p}_k$. In contrast, the ``line of sight'' emanating from the spacecraft in direction $\BH{d}_k$ will miss the star slightly due to errors in our measurement of $\BH{d}_k$, just as the spacecraft will not lie exactly on the line of position. Measuring the directions to two or more stars constrains the spacecraft to lie at the point where all the lines of position associated with the ensemble of stars intersect.

Since measurement errors prevent the lines of position from intersecting exactly, we are faced with the more complex task of describing the position and shape of the region of space where the spacecraft must be located to be consistent with all our direction measurements. \citet{kaplan} presented a solution for this problem, which served as a point of departure for the present analysis.  Figure \ref{fig:nav_wide} hints at this problem with dotted lines on either side of the star lines of position showing the transverse displacements of the lines if our direction measurements had differed by 1 arcsec.  Assuming the $\BV{p}_k$ are perfectly accurate, an angular uncertainty of $\sigma_\alpha$ in $\BH{d}_k$ causes a transverse  uncertainty $p_k\sigma_\alpha$ in the line of position, where $p_k= ||\BV{p}_k||$ is the distance to the star and $\sigma_\alpha$ is in radians.  This transverse uncertainty in spacecraft position can also be written $\sigma_\alpha/\pi_k$ in au, if both $\sigma_\alpha$ and the stellar parallax $\pi_k$ are in arcseconds. That is, the transverse displacement of each dotted line in au is numerically equal to the distance to the corresponding star in pc!  We usually think of 1 pc as the distance of a star for which a 1~au displacement of the observer changes its direction by $1''$, but for navigation problems, we think of it as the distance for which a $1''$ error in direction causes a 1~au transverse displacement of the line of position.  Of course, our actual angular uncertainties are not $1''.$ Figure \ref{fig:nav_wide} shows that we do significantly better; the spacecraft must actually be in a region with dimensions scaled by the ratio of our actual angle  uncertainties to $1''.$

The dotted lines in Figure \ref{fig:nav_wide} really represent the standard deviation of cylindrical Gaussian probability clouds surrounding lines P and W.  Our measurements constrain the spacecraft to the ellipsoidal Gaussian probability cloud that is the product of the individual cylindrical distributions.  The parallelogram where the dotted lines in Figure \ref{fig:nav_wide} cross shows the projected shape of this error ellipsoid (scaled up in size as if our angular  uncertainties were $1''$) --- just picture the ellipse inscribed in that parallelogram.  Given our measurements, the center of this error ellipsoid is the most likely position of the spacecraft, in other words, our best guess at $\BV{x},$ given the directions we measured. The shape and size of the error ellipsoid describe the uncertainty in the derived position $\BV{x}.$  We will now work out in detail how each additional direction measurement shrinks the position error ellipsoid in the two transverse directions. In brief, the task is to find the spacecraft position that minimizes (in a least-squares sense) the aggregate distances to all the lines of position.

If the true spacecraft position is $\BV{x},$ then the actual direction to star $k$ is $(\BV{x}-\BV{p}_k)/ ||\BV{x}-\BV{p}_k||.$ (These are the {\trl Gaia}+JPL entries in Table \ref{tab:coords}.)  Let $\delta\alpha_k$ be our measurement error --- the angle  on the sky between this actual direction and the measured direction $\BH{d}_k.$ Since $\delta\alpha_k$ and $ ||\BV{x}||/p_k$ are both very small (so the norm of the cross product of two unit vectors is the angle between them in radians),
\begin{equation} \delta\alpha_k = ||\BH{d}_k\CROSS(\BV{x}-\BV{p}_k)||/p_k,
\label{eq:dalphak}\end{equation}
which we rewrite as
\begin{equation} \delta\alpha_k^2 = \left( ||\BV{x}-\BV{p}_k||^2 - (\BH{d}_k\DOT(\BV{x}-\BV{p}_k))^2\right)/p_k^2.
\label{eq:alpha2}\end{equation}
If our measurements all have the same standard deviations in angle, a single direction measurement uncertainty $\sigma_\alpha,$ then each cylindrical Gaussian cloud will be $\exp(-\tfrac{1}{2}\delta\alpha_k^2/\sigma_\alpha^2),$ so the combined ellipsoidal cloud (that is, the probability density to measure $\BH{d}_0$ \textit{and} $\BH{d}_1$ \textit{etc.}) will be $\exp(-\tfrac{1}{2}\chi^2),$ where
\begin{equation} \chi^2 = \sum_k\delta\alpha_k^2/\sigma_\alpha^2.
\label{eq:chi2}\end{equation}
The center of this cloud is where $\chi^2$ is minimum, which is a perfectly well-defined point $\BV{x}$ even when the lines do not exactly meet (that is, when $\chi^2$ never reaches zero).

We thus use $\chi^2$-minimization to find the most likely parameters consistent with a set of measurements.  Here, $\chi^2$ is quadratic in $\BV{x},$ so setting its gradient to zero to find its minimum produces a simple $3\times 3$ system of linear equations for $\BV{x}.$ The gradient of eqn.(\ref{eq:alpha2}) is
\begin{align}
  \gradient\delta\alpha_k^2&= (2/p_k^2)\left((\BV{x}-\BV{p}_k)
                          - \BH{d}_k\BH{d}_k\DOT(\BV{x}-\BV{p}_k)\right) \nonumber\\
                        &= (2/p_k^2)\BV{Q}_k(\BV{x}-\BV{p}_k),
\label{eq:intermediate}\end{align}
where we have defined $\BV{Q}_k=\BV{I}-\BH{d}_k\BH{d}_k^T$ as the $3\times 3$ matrix which projects vectors into the plane perpendicular to $\BH{d}_k.$  We have also switched from vector notation with inner products written $\BV{a}\DOT\BV{b}$ to matrix notation with inner products written $\BV{a}^T\BV{b},$ and outer products $\BV{a}\BV{b}^T.$  Setting $ \gradient\chi^2=0$ and solving for $\BV{x}$ gives
\begin{equation} \BV{x} = \BL^{-1}\sum_k\BV{Q}_k\BV{p}_k/p_k^2,
\label{eq:x}\end{equation}
where
\begin{equation} \BL = \sum_k\BV{Q}_k/p_k^2.
\label{eq:xlambda}\end{equation}

Eqn.(\ref{eq:alpha2}) looks simpler when expressed in matrix notation using the $\BV{Q}_k$ projection operator, so we rewrite eqn.(\ref{eq:chi2}) as
\begin{equation} \chi^2 = (1/\sigma_\alpha^2) \sum_k(\BV{x}-\BV{p}_k)^T\BV{Q}_k(\BV{x}-\BV{p}_k)/p_k^2.
\label{eq:chi2a}\end{equation}
Now $\chi^2$ is just a quadratic in $\BV{x},$ and its second degree terms are proportional to $\BV{x}^T\left(\sum_k\BV{Q}_k/p_k^2\right)\BV{x},$ or $\BV{x}^T\BL \BV{x}.$ By construction, $\BL$ is a symmetric positive definite $3\times 3$ matrix, so its eigenvalues are positive and its eigenvectors form an orthogonal basis.  Since the ellipsoidal probability distribution for $\BV{x}$ is $\exp(-\tfrac{1}{2}\chi^2),$ those eigenvectors are the principal axes of the error ellipsoid \footnote{Specifically, the 1$\sigma$ error ellipsoid where $\chi^2=1.$  While the probability of falling within 1$\sigma$ error bars in 1D is roughly $2/3,$ the probability of falling within a 1$\sigma$ error ellipsoid in 3D is only roughly $1/5$ --- typically one of the three coordinates falls outside its 1D error bar!}, and the corresponding eigenvalues are the squares of its semi-axes.  Thus,
\begin{equation} \BV{V}_\BV{x} = \sigma_\alpha^2\BL^{-1} = \sigma_\alpha^2\left(\sum_k\BV{Q}_k/p_k^2\right)^{-1}
\label{eq:xcov}\end{equation}
is the covariance matrix for our most likely spacecraft position $\BV{x}$ given by eqn.(\ref{eq:x}).  Its eigenvalues (or singular values) are the squares of the principal axes of the error ellipsoid for the spacecraft position $\BV{x}$ calculated using eqn.(\ref{eq:x}).  Note that if $p_k$ are in parsecs and $\sigma_\alpha$ is in arcseconds, then the units of $\BV{V}_\BV{x}$ will be au$^2.$ Eqn.($\ref{eq:xcov}$) is a sort of matrix harmonic sum of the $p_k^2,$ with the $\BV{Q}_k$ projection matrices providing weighting to account for the angles among the stars.

Eqn.(\ref{eq:x}) matches the formulas given in \citet{kaplan}, with $1/p_k^2$ providing the weights for the terms in the sum.  That work focused on a series of observations separated in time by enough to determine both observer position and velocity vectors. Here the time separation of roughly one day is deliberately too short to be sensitive to motion of the spacecraft over the duration of the observations.  Given the velocity of the spacecraft and our astrometric precision, we would begin to be sensitive to its progression along its trajectory in about 12 days, which is an interval an order of magnitude longer than that used to obtain the present observations.  

Our $1/(p_k\sigma_\alpha)^2$ weighting assumes the angular uncertainties $\sigma_\alpha$ are the same in all directions.  Working out the more general forms for eqn.(\ref{eq:x}) and eqn.(\ref{eq:xcov}) for direction measurements with an arbitrary error ellipse on the sky is straightforward, changing the scalar $1/(p_k\sigma_\alpha)^2$ covariance into a $3\times 3$ matrix weight derived from the covariance matrix of the anisotropic  uncertainties.  The much simpler isotropic direction uncertainty formalism we present here captures all the important qualitative features of the more general result.

\subsection{Observational Errors}\label{sec:errors}

The accuracy of the New Horizon position determination directly depends on the angular astrometric  uncertainty $\sigma_\alpha.$  For a star at distance $p_k$ (in pc) an {\trl uncertainty} $\sigma_\alpha$ (in arcseconds) will cause the line determined by $\BH{d}_k$ and $\BV{p}_k$ to be displaced by $p_k\sigma_\alpha$ (in au) --- 1 pc is 1 au/arcsec! 
In practice, the astrometric quality of a given LORRI image depends on many factors, some of which can vary significantly over a sequence of exposures. As noted in $\S\ref{sec:nhobs}$, LORRI can provide accurate astrometry of given source in a program comprising several dozen images. The present demonstration was conducted with a small number of images, however, with the result that $\sigma_\alpha$ for Proxima Cen and Wolf 359 is only known to a factor of two.  One estimate of $\sigma_\alpha$ comes from the coordinate  uncertainties provided in Table \ref{tab:coords} for the New Horizons astrometric observations.  These are uncertainties in the mean, and are based solely on the observed random scatter among the six images used for each star. For a single LORRI image this measure implies $\sigma_\alpha=0\asec24$ averaged over both stars; however, the stars individually have considerably different uncertainties. Comparison with the {\trl Gaia}$+$JPL ``back of the book" coordinates give a similar value, but with the implied level of accuracy between the two stars reversed. It is likely that the limited number of observations prevents accurate determination of the true error distribution.

A different approach to estimate $\sigma_\alpha$ is to use the overall quality of the astrometric solution for each LORRI image, which is based on astrometric residuals of the few dozen stars in the LORRI fields surrounding the two target stars.  Over the ensemble of a dozen images, this yields a mean $\sigma_\alpha=0\asec 44.$ The uncertainty in $\sigma_\alpha$ argues for using the more conservative number, thus we take $\sigma_\alpha$ to be $0\asec 44$  for a single LORRI image in the present demonstration.   The pixel scale in LORRI $4\times4$-binned images is $4\asec08$, so this amounts to 0.11 pixel.  We note that \citet{lorri_nav} concluded that single LORRI measurements were good to 0.05 to 0.10 $4\times4$ pixels.   

At the same time we ignore any errors in the Gaia DR3 data, as they are orders of magnitude smaller than the LORRI astrometric errors. Gaia position errors are 20-60~$\mu$as at its baseline 2016.0 epoch.   Uncertainties in the Gaia proper motions rapidly degrade this accuracy over the 4.3~yr interval to April 2020, at least for high proper motion stars like Proxima Cen and Wolf 359. For them, the Gaia position  uncertainties have increased to of order 300 $\mu$as --- but that is still a factor of a thousand less than $\sigma_\alpha.$ Gaia parallax  uncertainties are also of order 50 $\mu$as causing an uncertainty in $p_k$ of about 20 au for Proxima Cen, and 80 au for Wolf 359.  These radial differences are projected transverse to the line of position from the star by the same tiny angle of a few seconds of arc shown in Figures \ref{fig:proxima} and \ref{fig:wolf}, again causing a completely negligible uncertainty in the transverse displacement of line P or W in Figure \ref{fig:nav_wide}.  (Changes in $p_k$ due to 4.3~yr of radial velocity are also tens of au, hence also negligible.)

Although the angular uncertainty in the direction measurement for a single LORRI image is about $0\asec44,$ we routinely reduce this uncertainty by averaging the measurements from several images of the same target.  In this work, we used six images each of Proxima Cen and Wolf 359, which should reduce $\sigma_\alpha$ by a factor of $\sqrt{6},$ to $0\asec 18$ or 0.044 pixel.  

We have aggregated our six images of each star into a single direction measurement for the purposes Figure \ref{fig:nav_zoom}, but we point out that eqn.(\ref{eq:x}) and eqn.(\ref{eq:xcov}) produce exactly the same values of both $\BV{x}$ and its covariance matrix (error ellipsoid) $\BV{V}_\BV{x}$ if one feeds the directions of all twelve images into them, as if they were $N=12$ individual stars.  That is, $\BV{x}$ for this $N=12$ set of directions will be exactly the same as if one had used the mean $\BH{d}_k$ from the six images of the same star in an $N=2$ application of eqn.(\ref{eq:x}).  Similarly, an $N=12$ version of Eq.(\ref{eq:xcov}) will produce the same error ellipsoid as the $N=2$ version with $\sigma_\alpha$ reduced by $\sqrt{6}.$  Taking each individual direction measurement as one term in these equations, so the number of terms equals the number of images rather than the number of separate targets, is the most convenient way to use them, and also provides crucial insights into how to design a navigation program.

\subsection{The Design of an Interstellar Navigation Program}

The size and shape of the error ellipse given by eqn.(\ref{eq:xcov}) can also be used as a design tool to develop an observational program for navigation.  Since the accuracy of the derived position is directly proportional to the distance of the star, only a few of the nearest stars are realistic candidates for this purpose.  Using eqn.(\ref{eq:xcov}), one can rank all of the pairs of these navigation candidate stars by their performance in constraining spacecraft position.  To do this, one sets $\BH{d}_k$ equal to $\BV{p}_k/p_k$ (a very good approximation), and does a singular value decomposition of the resulting $\BV{V}_\BV{x}/\sigma_\alpha^2$ matrix. The square roots of the largest and smallest singular values are the shortest and longest axes of the error ellipse for a position determination based on the directions of that pair of stars.  The best pair for determining spacecraft position will be the one with the smallest long-axis.  If one expresses $p_k$ in units of pc, the error ellipse axis will be in au per arcsecond of $\sigma_\alpha.$ Table \ref{tab:navstars} lists all five of the navigational candidate stars within 3~pc and the performance of the 10 pairs that can be constructed from them.

\begin{deluxetable}{lrrrrccccr}
\tabletypesize{\small}
\tablecolumns{10}
\tablewidth{0pt}
\tablecaption{Candidate Navigation Reference Stars}\label{tab:navstars}
\tablehead{
  \colhead{}&\colhead{}&\colhead{}&\colhead{}&\colhead{$d$}
  & \multicolumn{4}{c}{Long/Short Error Ellipsoid Axes (au/arcsec)} & \colhead{} \\
\multicolumn{1}{l}{Star} &
\multicolumn{1}{c}{$\alpha$} &
\multicolumn{1}{c}{$\delta$} &
\multicolumn{1}{c}{$m_V$} &
\multicolumn{1}{c}{(pc)} &
\multicolumn{1}{c}{B} &
\multicolumn{1}{c}{W} &
\multicolumn{1}{c}{H} &
\multicolumn{1}{c}{C} &
\multicolumn{1}{c}{SEA}
}
\startdata
Proxima Cen (P)   &217$^\circ$&$-63^\circ$& 11.1& 1.30& 1.91{/}1.06& 2.45{/}1.15& 2.70{/}1.16& 3.64{/}1.19& 117$^\circ$ \\
Barnard's Star (B)&269$^\circ$&  5$^\circ$&  9.5& 1.83&     & 2.58{/}1.46& 2.60{/}1.49& 6.54{/}1.56& 149$^\circ$ \\
Wolf 359 (W)      &164$^\circ$&  7$^\circ$& 13.5& 2.41&     &     & 7.00{/}1.75& 3.80{/}1.87&  56$^\circ$ \\
HD 95735 (H)      &166$^\circ$& 36$^\circ$&  7.5& 2.55&     &     &     & 4.24{/}1.93&  53$^\circ$ \\
 CD-23 14742 (C)   &282$^\circ$&$-24^\circ$& 10.4& 2.98&     &     &     &     & 174$^\circ$ \\
\enddata
\tablecomments{Candidate navigation stars within 3 pc ($\alpha$ Cen, Sirius, and UV Ceti, excluded because they are multiple star systems).  The error ellipsoid axes columns give the longest{/}shortest axes of the error ellipsoid computed using Eq.(\ref{eq:xcov}) for the corresponding pair; au/as is position  uncertainty in au per angular  uncertainty in arcseconds, which is an effective distance in pc.  The SEA column gives the solar elongation angle of the star as seen from NH (see $\S\ref{sec:parallax}$).}
\end{deluxetable}

Careful study of Table \ref{tab:navstars} clarifies eqn.(\ref{eq:xcov}).  The distance column can be read as if its units were au/arcsec rather than pc --- how much measuring the direction to that star will constrain the transverse spacecraft position per arcsecond of angular uncertainty.  Continuing across a row, the entries show how much measuring a second star improves the constraint on spacecraft position: The first number is the longest axis of the error ellipsoid for the pair, while the second number is the shortest.  With only one star, the spacecraft position was completely unconstrained in the direction along the line of position; adding a second star trims this error cylinder to a finite error ellipse.  If the directions of the two stars were at right angles, the long axis (in au/arcsec) would be exactly the same as the distance to the farther star (in pc).  As the angle decreases, the second star fails to constrain the position as tightly, as is evident for the W and H pair.   In the limit that the two stars are 0 or 180 degrees apart, the long axis of the error ellipsoid becomes infinite and the pair is not useful.  On the other hand, the shortest axis of the error ellipsoid is always in the direction perpendicular to both lines of position (the dashed line in Figure~\ref{fig:nav_zoom}).  The inverse square of this short axis of the error ellipsoid is the sum of the inverse squares of the two distances (e.g.- $1/1.15^2 = 1/1.30^2 + 1/2.41^2$), always smaller than the distance to either star, and independent of the angle between them.

Proxima Cen and Barnard's Star, the two nearest stars, are by a considerable margin the best choice for interstellar navigation, at least for a spacecraft anywhere within tens of thousands of au of the Sun.  Our choice of Proxima Cen and Wolf 359 is the second best pair, with about 30\% lower position accuracy --- 2.45 versus 1.91~au/arcsec.  Using Barnard's Star and Wolf 359 is only slightly worse at 2.58~au/arcsec, if Proxima Cen cannot be used.  (But given their ecliptic latitudes, a spacecraft is far more likely to be unable to view Wolf 359 than either Proxima Cen or Barnard's Star.)  We rejected Barnard's Star for this work because the sine of its SEA is small enough that its shift in position on the sky between observers on Earth and New Horizons is only half of that realized with Wolf 359.  However, we point out that the SEA angle makes no difference at all if the goal is to minimize the uncertainty in derived position.  Using Barnard's Star instead of Wolf 359 would have reduced our uncertainty in $\BV{x}$ by nearly 30\%. In other words, a nearby star with a small ES parallax nevertheless provides an accurately determined line of position that can be used in combination with a suitably positioned second nearby star.

We originally thought that increasing the number of {\it different} stars observed would be the most efficient way to improve the accuracy of the spacecraft position.  However, eqn.(\ref{eq:xcov}) tells a different story.  One can use it to assess the performance of using any set of stars by the size of the resultant error ellipse.  Doing this, we discovered that it is almost always more efficient to repeat an image of a closer star that has already been used, rather than to add an image of a third more distant star.  This is certainly true of Proxima Cen and Wolf 359: If one wants to improve the accuracy of $\BV{x},$ the best option is to reimage Wolf 359 to tighten its astrometric uncertainties, rather than using the same resources to observe any more distant star. In short, for navigation, the optimal strategy is to choose the best pair of stars possible, and image them to the degree needed to beat down their effective $\sigma_\alpha.$   Eventually, systematic errors limit how much repeated imaging of the same star can reduce its position uncertainty.  To the extent that these systematic errors vary randomly from star to star, one can imagine scenarios in which additional stars could improve navigation accuracy.  However, we believe that in practice the best way to navigate is to concentrate on the nearest two or at most three stars.

It is a remarkable accident that the directions to the three nearest candidate navigation stars very nearly form an orthonormal basis.  (The P-B, P-W, and B-W angles are 78$^\circ$, 81$^\circ$, and 104$^\circ$ degrees.)  Studying the case of three stars with mutually orthogonal directions and unequal distances provides considerable insight into navigation program design.  For example, suppose the distances $p_0$ and $p_1$ are both $\sqrt{2}$~pc, and the distance $p_2$ is 2~pc.  With two images each of the first two stars, the position error ellipsoid will be 1~au (per arcsecond direction  uncertainty) in the $\BH{d}_0$ and $\BH{d}_1$ directions, and $1/\sqrt{2}$~au in the $\BH{d}_2$ direction.  On the other hand, with only one image each of the first two stars, and two images of the third star, the error ellipsoid will be a 1~au sphere.  In other words, with these three stars, there are two different ways to achieve 1~au position accuracy in every direction with a total of four images.  If the distance $p_2$ to the third star were any larger, there is no benefit to imaging it - an example of the rule that sticking with a pair of navigation stars is the best option.  On the other hand, if $p_2<2$~pc then the maximum dimension of the error ellipsoid decreases by trading one image of each of the first two stars for two images of third.  The general rule is that it is pointless to consider imaging a third star if $p_2>(p_0^2+p_1^2)^{1/2},$ while if the third star is closer than this, there may be a benefit to imaging it.  If the first two stars are Proxima Cen and Barnard's Star, $(p_0^2+p_1^2)^{1/2}=2.24$~pc, and no third star is close enough to be worth imaging.

\subsection{The Derived Spatial Location of New Horizons}

\begin{figure}[hbtp]
\centering
\includegraphics[keepaspectratio,width=7.0 in]{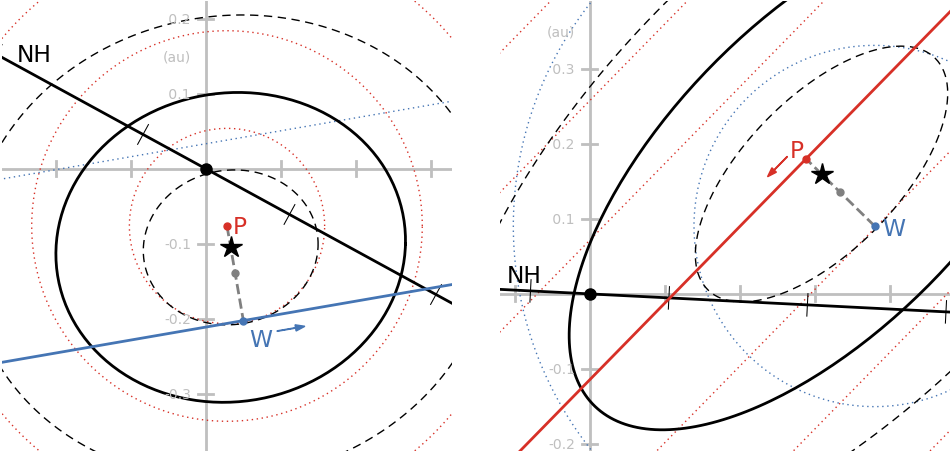}
\caption{The lines of position for Proxima Cen (P) and Wolf 359 (W) covering the area of space where they both pass by New Horizons, viewed from two directions. The small arrows near the P and W labels indicate the direction to the given star. The left view looks in the observed direction to Proxima Cen; the right view is the same for Wolf 359.  NH marks the true trajectory of New Horizons. Tick marks are at 30 day intervals; the spacecraft moves right to left in both views.  The gray axes with ticks at 0.1~au intervals intersect at the true position of NH on 2020 April~23.  The ticks bracketing this point are April~5 and May~5, respectively.  The P and W lines do not intersect; the points of closest approach of the two lines are connected by the dashed gray line, which lies in the plane in both views (positive ecliptic latitude is up in both views).  The P and W lines are 0.133~au apart at closest approach.  The midpoint of the dashed connecting line is indicated by a gray dot.  The most likely spacecraft position $\BV{x}$ from eqn.(\ref{eq:x}) is indicated by the black star.  It is 0.351~au from the true spacecraft position indicated by the black dot.  The faint dotted lines are nested cylinders around the P and W lines corresponding to steps of $0\asec 1$ in observed direction. The solid ellipses are the 1$\sigma$ error ellipses from eqn.(\ref{eq:xcov}) assuming $\sigma_\alpha=0\asec 44$ for each individual LORRI image; the dashed ellipses are 0.5$\sigma$ and 1.5$\sigma$.  The spacecraft is at  0.94$\sigma.$}
\label{fig:nav_zoom}
\end{figure}

Figures \ref{fig:nav_wide} and \ref{fig:nav_zoom} present our navigation solution for New Horizons on 2020 April 23 with respect to the orbits of the outer planets and the vicinity of the true location of the spacecraft, respectively.  The solution is also presented quantitatively, and compared to the true spacecraft position, in Table \ref{tab:nav}.

The highly zoomed-in view presented in Figure \ref{fig:nav_zoom} provides a detailed graphical exposition of how the directions to the reference stars observed by New Horizons are combined to estimate the position of the spacecraft.
The left panel is a zoomed-in view of the same geometry depicted in Figure \ref{fig:nav_wide}, viewed in the measured direction of Proxima Cen, so that  line P projects to a point.  The right panel is the analogous view in the direction of Wolf 359, projecting line W to a point.  These are aggregated lines: the directions $\BH{d}_k$ are the means over the six images of each star.  The faint dotted circles and lines correspond to steps of $0\asec 1$ away from the nominal directions defining the P and W lines; recall that the estimated $\sigma_\alpha$ is $0\asec 44/\sqrt{6}=0\asec 18.$ The star-symbol marks our best estimate of the spacecraft position $\BV{x}$ from eqn.(\ref{eq:x}).  The actual spacecraft position on 2020 April 23 according to JPL Horizons is at the intersection of the two gray axes.  It is  0.351~au away from the starred position derived from our LORRI images. The solid ellipse is the 1-$\sigma$ error ellipse from eqn.(\ref{eq:xcov}).  The actual position is just inside this ellipse (at  0.94~$\sigma$), consistent with our assumed $0\asec 44$  uncertainty for a single LORRI image.  {\dhm The principal semi-axes of the error ellipsoid are 0.44, 0.23, and 0.21 au, with the long axis roughly in the direction of Proxima Cen, as shown in Figure \ref{fig:nav_zoom}.}

\begin{deluxetable}{lllllll}
\tabletypesize{\small}
\tablecolumns{7}
\tablewidth{0pt}
\tablecaption{The Position of New Horizons}
\tablehead{
\multicolumn{1}{c}{} &
\multicolumn{1}{c}{$\alpha$} &
\multicolumn{1}{c}{$\delta$} &
\multicolumn{1}{c}{Range (au)} &
\multicolumn{1}{c}{x (au)} &
\multicolumn{1}{c}{y (au)} &
\multicolumn{1}{c}{z (au)}}
\startdata
NH (LORRI)& $288\adeg 11(38)$& $-20\adeg 21(45)$& $\efill46.89(27)$& $\efill13.68(28)$& $-41.82(23)$& $-16.20(40)$ \\
JPL mean& $287\adeg 8723\phantom{0}$& $-20\adeg 4433$& $\efill47.1176$&  $\efill13.5495$& $-42.0195$&  $-16.4573$ \\
JPL $\Delta$& $\efill\pm0\adeg 0006$& $\efild\pm0\adeg 0001$& $\pm0.0046$& $\pm0.0018$& $\efild\mp0.0040$& $\efild\mp0.0015$ \\
NH$-$JPL& $\phantom{00}0\adeg 24$& $\phantom{-0}0\adeg 23$& $-0.23$& $\efill\phantom{0}0.13$& $\phantom{-0}0.20$& $\phantom{-0}0.26$\\
\enddata
\tablecomments{The first row tabulates the New Horizons position vector derived solely from LORRI images and the Gaia star catalog \citep{gaia}.  Standard  deviations assuming  single image $\sigma_\alpha$ is 0\asec 44 are indicated by the two digits in parentheses. The JPL mean row gives the average positions, based on NASA's Deep Space Network tracking, over the 27.65 hour period in which all twelve LORRI images were made.  The JPL $\Delta$ row shows how much the spacecraft moved in those 27.65 hours.  The NH$-$JPL row shows the actual errors of our navigation method, for comparison to the estimated uncertainties in the first row.}

\end{deluxetable}\label{tab:nav}

For the case of two lines, it is clear that the minimum $\chi^2$ must occur on the shortest line segment connecting the two lines, which is the dashed gray line in Figure \ref{fig:nav_zoom}, because one can decrease the $\chi^2$ of any point off this line by zeroing its component perpendicular to the line.  This dashed line lies in the planes of both the left and right views; its direction is the cross product of the directions of the P and W lines.  The most likely point (the starred point) divides this connecting line into two parts with lengths in the ratio $p_1^2:p_2^2.$ Note that as the more distant star gets farther and farther away, the $\BV{x}$ from eqn.(\ref{eq:x}) rapidly converges on the line determined by the nearer star; at the same time, the error ellipse from eqn.(\ref{eq:xcov}) becomes more and more elongated about the nearer-star line.  This tendency is already evident in Figure \ref{fig:nav_zoom}, given that Wolf 359 is $1.85\times$ farther away than Proxima Cen.

Table \ref{tab:nav} directly compares our derived celestial coordinates and range to New Horizons (referenced to the SSB) to the true values provided from the NASA/JPL Horizons Solar System ephemeris.  Detailed understanding of the  uncertainties in any component of the position vector must reflect the three-dimensional error ellipsoid presented in Figure \ref{fig:nav_zoom}. But in broad detail, our celestial coordinate estimate was within a quarter of a degree of the true direction in both right ascension and declination. To make this more concrete, a quarter of a degree is half the angular diameter of the full Moon.  The range error is $-0.23$~au, while the overall error is $0.35$~au. As pedagogical exercise, we note that this error is no more than the width of the trajectory line plotted on the solar system map shown in Figure \ref{fig:nav_wide}.

\section{The Utility of Stellar Astrometry For Interstellar Navigation}

The accuracy of the ES parallaxes that we obtained is low.  It does not approach that of standard Earth-based parallax programs, let alone the spectacular accuracy delivered by the Gaia DR3 catalog \citep{gaia}.  The accuracy that we obtained for the components of the New Horizons position vector on 2020 April 23 is more interesting, but we likewise do not begin to approach the navigational tracking abilities of the DSN.  This is not surprising. LORRI uses a small telescope that feeds a CCD with large pixels, and the program obtained only minimal observations of the two reference stars.  

That said, we achieved the goals that we set out to.  The parallax program provides a simple and instantaneous demonstration of parallactic observations, free from the effects of proper motion, and with shifts so large that they are obvious by simple visual inspection of the paired Earth and spacecraft images. The derived NH position vector is accurate enough to provide a credible location of the spacecraft in the outer Solar System. This represents the first time that optical stellar astrometry has been used to determine the three-dimensional location of a spacecraft with respect to stars in the local solar neighborhood. It is also the first time any method of interstellar navigation
has been demonstrated for a spacecraft on an  interstellar trajectory. Indeed, the use of X-ray pulsars for navigation has only been demonstrated in low earth orbit.

An obvious question is to what extent we can improve the accuracy of the present navigation technique.  As noted in the previous section, analysis shows that the most efficient enhancement would be to improve the astrometric accuracy of the direction determinations for the single pair of stars that provides the most compact error ellipsoids, as opposed to simply observing a larger sample of reference stars. For any future navigation demonstrations using New Horizons, quadrupling the number of images obtained over the present program should be feasible, with the goal of reducing the random  uncertainties in the mean positions of the stars by $2\times$. According to the analysis in Table \ref{tab:navstars}, switching out Wolf 359 for Barnard's star would effect an additional uncertainty reduction of $\sim20$\%.  With an input uncertainty of $\sigma_\alpha=0\asec18,$ this would reduce the long axis of the final error ellipsoid to 0.17 au.  We reiterate the interesting conclusion that even though Barnard's star has a relatively small ES parallax, the Proxima Cen/Barnard's star pair provides the best navigational reference for the first leg of a voyage departing the Solar System for interstellar space.

Considerably better performance should be possible using the cameras presently deployed on other interplanetary spacecraft, or contemplated for future missions. Telescopes with apertures plausibly larger than LORRI's, with diffraction-limited optics, delivering images to Nyquist-sampled detectors, mounted on platforms with matching fine-pointing control, should be able to provide astrometry with few milli-arcsecond accuracy. Extrapolating from LORRI, position vectors with accuracy of 0.01 au should be possible in the near future.

Is this interesting? These crude estimates are still orders of magnitude short of what can be done with DSN distance-ranging and $\Delta$-DOR (Delta-Differential One-Way Ranging; see \citealt{lorri_nav}) for the directional location of the spacecraft. Further, if the concern is the ability to do {\it autonomous} navigation from the spacecraft alone, then the pulsar-based methods cited at the top of the previous section will provide the best approach for spacecraft that can be easily equipped with the needed instrumentation.  On the other hand, being able to navigate with a camera primarily designed for science objectives is likely to be an economical solution, and 0.01 au range accuracy may be interesting in some contexts.  We also note that spacecraft are being designed with ever increasing levels of autonomous operation.  The image processing and analysis required to support the present demonstration was not particularly challenging and is well within the capabilities of modern flight systems. Data down-link volume was a strongly limiting resource for New Horizons, but would not be a factor with on-board analysis.

Using the stars to navigate is an ancient technique.  Most inter-planetary spacecraft carry instrumentation that can do celestial navigation at some level, and high resolution imaging cameras in general should be well suited to this task.  Our simple demonstration was done with a small camera developed primarily to image Pluto. It is now twice as far away from the Sun, the most remote camera ever operated.  The more we step out in space the more the familiar positions of the stars change.  The present demonstration documents this, and shows that any spacecraft leaving Earth-space for destinations in the Solar System or beyond can document their own travels as well.

\acknowledgments

We dedicate this work to the memory of Chad K\=alepa Baybayan, a master Pwo navigator, who dedicated his life to the teaching and demonstration of the ancient Polynesian science of maritime and celestial navigation over the vast reaches of the Pacific ocean and beyond. It was our pleasure to have discussed this work with him during its initial definition. We thank NASA for funding and continued support of the New Horizons mission, which were required to obtain the present observations.  We thank the referee for a careful reading of the manuscript and suggestions for improving the clarity  of the narrative. The data presented were obtained during the second Kuiper Extended Mission of New Horizons. TRL is funded by NSF NOIRLab, which is managed by AURA under a cooperative agreement with the National Science Foundation. This work made use of data from the European Space Agency (ESA) mission {\it Gaia} (\url{https://www.cosmos.esa.int/gaia}), processed by the {\it Gaia} Data Processing and Analysis Consortium (DPAC, \url{https://www.cosmos.esa.int/web/gaia/dpac/consortium}). 

\software{matplotlib \citep{matplotlib}, Vista \citep{vista}}

{}

\end{document}